\documentclass[pra,aps,twocolumn,showpacs]{revtex4}

\usepackage{graphicx}
\begin{document}
\def\pd#1#2{{\partial #1\over\partial #2}}

\title{\bf Scenario of strongly non-equilibrium Bose-Einstein condensation}
\author {Natalia G. Berloff${}^1$ and Boris V. Svistunov${}^2$}
\affiliation{ 
${}^1$Department of Mathematics, University of California, Los
Angeles, CA, USA, 90095-1555\\ ${}^2$ Russian Research Center
``Kurchatov Institute'', 123182 Moscow, Russia}
\email{
nberloff@math.ucla.edu,  svist@kurm.polyn.kiae.su}
\date {July 10, 2001}
\begin {abstract}
Large scale numerical simulations of the Gross-Pitaevskii equation
are used to elucidate the self-evolution of a Bose gas from  a
strongly non-equilibrium initial state. The  stages of the process
confirm and refine the theoretical scenario of Bose-Einstein
condensation developed by Svistunov, Kagan, and Shlyapnikov
\cite{svist,kss,ks1}: the system evolves from the regime  of weak
turbulence to superfluid turbulence via states  of strong
turbulence in the long-wavelength  region of  energy space.
  \end{abstract}
\maketitle

\section{Introduction}
\subsection{Statement of the problem}
\label{a}
The experimental realization of Bose-Einstein condensates (BEC) in
dilute alkali and hydrogen gases \cite{bec1}  and more recently in
a gas of metastable helium \cite{bec2} has stimulated a great
interest in the dynamics of BEC. In the case of a pure condensate,
both equilibrium and dynamical properties of the system can be
described by the Gross-Pitaevskii equation (GPE)\cite{gp} (in
nonlinear physics this equation is known as defocusing nonlinear
Schr\"odinger equation). The GPE has been remarkably successful in
predicting the condensate shape in an external potential, the
dynamics of the expanding condensate cloud, the motion of
quantized vortices; it is also a popular qualitative model of
superfluid helium.

An important and often overlooked feature of the GPE is that it
gives an accurate microscopic description of the formation of BEC
from the strongly degenerate gas of weakly interacting bosons
\cite{ly,ks2}. By  large scale numerical simulations of the GPE it is
possible, in principle, to reveal all the stages of this evolution
from weak turbulence to superfluid turbulence with a tangle of
quantized vortices as was argued by Svistunov, Kagan, and
Shlyapnikov \cite{svist,kss,ks1} (for a brief review see
Ref.~\onlinecite{svist2}). This task has up to now remained
unfulfilled, though some important steps in this direction have
been  made in Refs.~\cite{Damle,Drummond}. We would also like to mention
 the description of the equilibrium fluctuations of the condensate {\it
and} highly occupied {\it non-condensate} modes
 using the  time-dependent
GPE \cite{Goral,Davis}.

The goal of this paper is to obtain  the  conclusive description of
the process of strongly non-equilibrium BEC formation in a
macroscopically large uniform weakly interacting Bose gas using
the GPE. We are especially interested in tracing the development
of the so-called coherent regime \cite{ly,svist,kss,ks1} at a
certain stage of evolution. According to the theoretical
predictions \cite{kss,ks1},  this regime sets in after the
breakdown of the regime of weak turbulence in a low-energy region
of wavenumber space. It  corresponds to the formation of the
superfluid short-range order which is the state of superfluid
turbulence with quasi-condensate local correlation properties.

 The
notions of weak turbulence and superfluid turbulence are  crucial
to  our understanding of ordering kinetics. In the regime of weak
turbulence (for an introduction to the  weak turbulence theory for
the GPE see Ref.~\onlinecite{Nazarenko}), the ``single-particle"
modes of the field are almost independent due to weak nonlinearity
of the system. The smallness of correlations between  harmonics in
the regime of weak turbulence implies the absence of any order. On
the other hand, 
 the regime of superfluid turbulence (for an introduction, see
Ref.~\onlinecite{Donnelly}) is the regime of strong coherence
where the local correlation properties correspond  to the
superfluid state, but the long-range order is absent because of
the presence of a chaotic vortex tangle and non-equilibrium
long-wave phonons \cite{ks1}. In the case of a weakly interacting,
gas  the local superfluid order is synonymous to the existence of
quasi-condensate correlation properties \cite{kss}. In a
macroscopically large system, the cross-over from weak turbulence
to superfluid turbulence is a key ordering process. Indeed, in the
regime of weak turbulence there is no order at all, while in the
regime of superfluid turbulence the (local) superfluid order has already been
formed.  Meanwhile, a rigorous theoretical as well as
numerical or experimental studies of this stage of evolution have
been lacking. The general conclusions concerning this stage
\cite{kss,ks1} were made on the basis of qualitative analysis
that naturally contained {\it ad hoc} elements. The difficulty
with an accurate analysis of the transition  from weak turbulence
to  superfluid turbulence comes from the fact that the
evolution between these two qualitatively different states takes place
 in the regime of strong turbulence which is hardly amenable to
analytical treatment. Large
computational resources are necessary for a numerical analysis of this stage since the problem involves
significantly different length scales and, therefore, requires high
spatial resolution.

In the present paper we demonstrate that this problem can be
unambiguously solved with a powerful enough computer. Our numerics
clearly reveal the dramatic process of transformation from weak
turbulence to  superfluid turbulence and, thus,  fills in a
 serious gap in rigorous theoretic description of the strongly
non-equilibrium BEC formation kinetics in a macroscopic system.

The paper is organized as follows. In Sec.~\ref{b} we discuss the
relevance of the time-dependent GPE to the description of the BEC
formation kinetics and its relation to the other formalisms. In
Sec.~\ref{c} we render some important details of the evolution
scenario that we are going to observe. In Sec.~\ref{sec2} we
describe our numerical procedure. In Sec.~\ref{sec3} we present the
results of our simulations. In Sec.~\ref{sec4} we conclude with
outlining the observed evolution scenario and making a comment on
the case of a confined gas.

\subsection{Time-dependent Gross-Pitaevskii equation and BEC formation kinetics}
\label{b}

In this Section
we will  discuss  the question
of applicability of the GPE to the BEC formation kinetics, and its
connections to the other---full-quantum---treatments. These discussion is especially relevant in the wake of a recent controversy on
the applicability of the classical-field description
to a non-condensed bosonic field.

A general analysis of the kinetics of a weakly interacting bosonic
field was performed in Ref.~\onlinecite{cd}. In terms of the
coherent-state formalism, it was demonstrated that if the occupation
numbers are large and somewhat uncertain (with the absolute value of
the uncertainty being much larger than unity and with a
relative value of the uncertainty being arbitrarily small), then
 the system evolves as an ensemble
of classical fields with corresponding classical-field action.
(For an elementary demonstration of this fact for a
weakly interacting Bose gas and especially for the discussion of
the structure of the {\it initial state} see
Ref.~\onlinecite{ks2}.) This  has a direct analogy with the
electro-magnetic field: (i)~the density matrix of a completely
disordered  weakly interacting Bose gas with large
and somewhat uncertain  occupation
numbers is almost diagonal in the coherent-state representation,
so that the initial state can be viewed as a mixture
or statistical ensemble of coherent states; (ii)~to the leading order
 each coherent state evolves along its
classical trajectory which in our case is given by the GPE
\begin{equation}
{\rm i}\hbar \pd \psi t = -{\frac{\hbar^2}{2m}} \nabla^2 \psi + U|\psi|^2 \psi,
\label{gpDim}
\end{equation}
where $\psi$ is the complex-valued classical field that specifies
the index of the coherent state, $m$ is the mass of the boson,
$U=4\pi\hbar^2 a/m$ is the strength of the $\delta$-function
interaction (pseudo-)potential, and  $a$ is the scattering length.
[Note that in a strongly interacting system it is
impossible to divide  single-particle modes into highly
occupied and practically empty ones, so the requirement of weak interactions is essential here. In a strongly interacting
system, there are always  quantum modes with occupation
numbers of order unity that are  coupled to the rest
of the system.] Therefore, the
behavior of the quantum field is equivalent to that of an ensemble
of classical matter fields.

It is important to emphasize  that in the context of the strongly
non-equilibrium BEC formation kinetics the condition of large
occupation numbers is self-consistent: the evolution leads to an
explosive increase of occupation numbers in the low-energy region
of wavenumber space \cite{svist} where the ordering process takes
place.  Even if the occupation numbers are of order unity in the
initial state, so that the classical matter field description is
not yet applicable, the evolution that can be described at this
stage by the standard Boltzmann quantum kinetic equation
inevitably results in the appearance of large occupation numbers
in the low-energy region of the particle distribution (see, e.g.,
Ref.~\cite{semikoz}). The blow-up scenario \cite{svist} indicates
that only  low-energy part of the field is initially involved in
the process. Therefore, one can  switch from the kinetic equation
to the matter field description for the long-wavelength component
of the field at a certain moment in the evolution when the
occupation numbers become appropriately large. As the time scale
of the formation of the local quasi-condensate correlations is
much smaller than any other characteristic time scale of evolution
\cite{kss}, the cut-off of the high-frequency modes, associated
with the matter field description, is not important. By the time
the interactions (particle exchange) between the high- and
low-frequency modes became significant, the local superfluid order
had already been developed. The order of interaction wavelengths is of
typical thermal deBroglier wavelength and, therefore,  these
interactions are  essentially local with respect to the
quasi-condensate and can be described in terms of the kinetic
equation\cite{svist,semikoz}.

The thesis of the applicability of the matter field description at
large occupation numbers was justified by the analysis of
Ref.~\onlinecite{cd}. Later  Stoof questioned the validity of
this thesis  by introducing the concept of ``quantum nucleation"
of condensate as a result of an essentially quantum instability
\cite{Stoof}; the path-integral version of Keldysh
formalism was used to substantiate this concept. For a criticism of the
concept of ``quantum nucleation'' see
Refs.~\onlinecite{ks2,commentStoof}.

It is important to emphasize, however, that the path-integral
approach developed in \cite{Stoof} appears to be the most
fundamental, powerful, and universal way of deriving the basic
equations for the dynamics of a weakly interacting Bose gas. In
particular, we believe that  the demonstration of the applicability
of the time-dependent GPE to the description of highly occupied
single-particle modes of a non-condensed gas within this formalism
would be the most natural since the effective action for the
bosonic field is simply the classical-field action of the GPE.
Basically, one simply has to make sure that for the modes with
large and somewhat uncertain occupation numbers the main
contribution to the path integral comes from a close vicinity of
the classical trajectories with the quantum corrections being
relevant only on large enough times of evolution.

An interesting all-quantum description of the BEC kinetics was
implemented in Ref.~\onlinecite{Drummond}. This technique is based
on associating the quantum-field density matrix in coherent-state
representation with a correlator of a pair of classical fields
which evolution is governed by a system of two coupled nonlinear
equations with stochastic terms. Using this method the authors
performed a numerical simulation of the BEC formation in a trapped
gas of a moderate size. We believe (in particular, in view of
general results of Refs.~\onlinecite{cd,ks2}) that this approach
might be further developed {\it analytically} to demonstrate
explicit overlapping with the other treatments and 
with the time-dependent GPE. Indeed, the form of the system of two coupled
 equations of Ref.~\onlinecite{Drummond} is
reminiscent of that of the GPE. This suggests that under the
condition of large occupation numbers the system can be decoupled
 leading to the GPE for the diagonal part of the density matrix with
 relative smallness of the
non-diagonal terms. If the
standard Boltzmann equation is applicable, so that  the system can be viewed as
an ensemble of weakly
coupled elementary modes \cite{Gardiner}, it is natural to expect that the
equations of Ref.~\onlinecite{Drummond} should lead to the kinetic
equation. A natural way for deriving kinetic
equation from the dynamic equations of
Ref.~\onlinecite{Drummond} is to utilize the standard formalism
of the weak turbulence theory. In the case of the GPE, the weak
turbulence approximation leads to the quantum-field Boltzmann
kinetic equation without {\it spontaneous} scattering processes
(see, e.g., Ref.\cite{Nazarenko}; note also that it is the
simplest way to make sure that the GPE is immediately applicable once
the occupation numbers are large). It is natural to expect that
in the full-quantum treatment of Ref.~\onlinecite{Drummond} the
weak-turbulence procedure over the dynamic equations would result
in the complete quantum-field Boltzmann kinetic equation with the
spontaneous processes retained. Unfortunately, we are not aware of
such investigations of the equations of
Ref.~\onlinecite{Drummond} that might be very instructive for
the general understanding of the dynamics of a weakly interacting
Bose gas.

\subsection{Initial state and evolution scenario}
\label{c}

In what follows we consider the evolution of Eq.~(\ref{gpDim})
starting with a strongly non-equilibrium initial condition
\begin{equation}
\psi({\bf r},t=0)=\sum_{\bf k} a_{\bf k} \exp ({\rm i}{\bf kr}),
\label{incond0}
\end{equation}
where the phases of the complex  amplitudes $a_{\bf k}$ are
distributed randomly. Such an initial condition follows from the
microscopic quantum-mechanical analysis of the state of a weakly
interacting Bose gas in the kinetic regime \cite{ks2}. Theoretical
investigations of the relaxation of such an initial state towards
the equilibrium configuration were performed by Svistunov, Kagan,
and Shlyapnikov \cite{svist,kss,ks1}. The analysis revealed a
number of stages in the  evolution. Initially  the system is in
the weak turbulence regime and thus can be described by Boltzmann
kinetic equation. The kinetic equation is obtained as the
random-phase approximation of Eq.~(\ref{gpDim}) for occupation numbers
$n_{\bf k}$ defined by $\langle a_{\bf k} a_{\bf k'}^*\rangle
\approx n_{\bf k} \delta_{\bf k}\delta_{\bf k'}$. Alternatively,
the weak-turbulence kinetic equation follows from the general
quantum Boltzmann kinetic equation if one neglects  spontaneous
scattering as compared with   stimulated scattering (because of
large occupation numbers). Svistunov \cite{svist} and later
Semikoz and Tkachev \cite{semikoz} considered the self-similar
solution of the Boltzmann kinetic equation:
\begin{eqnarray}
n_\epsilon(t)&=& A \epsilon_0^{-\alpha}(t)f(\epsilon/\epsilon_0),
\qquad t\le t_*,\label{k1}\\
\epsilon_0(t)&=&B(t_*-t)^{1/2(\alpha-1)}, \label{k2}\\
f(x)&\rightarrow&x^{-\alpha} \quad {\rm at} \quad x\rightarrow
\infty, \quad f(0)=1, \label{k3}
\end{eqnarray}
where  $\epsilon=\hbar^2 k^2/2m$. The dimensional constants  $A$
and $B$ relate to each other by $(\alpha-1)m^3U^2A^2 = \lambda
\pi^3 \hbar^7 B^{2(\alpha-1)}$, where the parameters $\alpha$ and
$\lambda$ were determined by numerical analysis as $\alpha \approx
1.24$ \cite{semikoz} and $\lambda\approx 1.2$ \cite{svist}. The
form of the function $f$ was also determined numerically in Ref.~\cite{svist}.
 The solution (\ref{k1})-(\ref{k3}) has only one free
parameter, say $A$, that depends on  conditions in the
``pre-historic" evolution. By ``pre-historic'' evolution we mean any
sort of non-universal dynamics preceding the appearance of
self-similarity. This dynamics is sensitive to the details of the
initial condition or/and cooling mechanism as well as to the
spontaneous-scattering terms in the kinetic equation which cannot
be neglected until the occupation numbers are large enough. 
When  the
self-similar regime sets in at a certain step of evolution all the particular details of the
previous evolution are absorbed in the single parameter $A$.

The self-similar solution  (\ref{k1})-(\ref{k3}) describes a wave
in  energy space propagating from high to lower energies. The
energy $\epsilon_0(t)$ defines  the ``head" of the wave. The wave
propagates in a blow-up fashion: $\epsilon_0(t) \rightarrow 0$ and
$n_{\epsilon_0}(t)\rightarrow \infty$ as $t \rightarrow t_*$. In
reality the validity of the kinetic equations associated with
the random phase approximation breaks down shortly before the
blow-up time $t_*$. This moment marks the beginning of a
qualitatively different stage in  the evolution the coherent
regime:  strong turbulence evolves into a quasi-condensate
state. In the coherent regime the phases of the complex
amplitudes $a_{\bf k}$ of the field $\psi$ become strongly
correlated and the periods of their oscillations are then
comparable with the evolution times of the occupation numbers. The
formation of the quasi-condensate is manifested by the appearance
of a well-defined tangle of quantized vortices and, therefore, by the
beginning of the final stage of the evolution: superfluid
turbulence. In this regime the vortex tangle starts to relax over
macroscopically large times.
\section{Numeric procedure}
\label{sec2}

\subsection{Finite-difference scheme}
\label{sec2:a}

 We  performed  a large scale numerical integration of a
dimensionless form of the GPE
\begin{equation}
-2{\rm i} \pd \psi t =  \nabla^2 \psi + |\psi|^2 \psi,
\label{gpDimless}
\end{equation}
starting with a strongly non-equilibrium initial condition. Our
calculations were done in a periodic box $N^3$, with $N=256$,
using a fourth-order (with respect to the spatial variables)
finite-difference scheme. The scheme corresponds to the
Hamiltonian system in the discrete variables $\psi_{ijk}$:
\begin{equation}
{\rm i}\pd {\psi_{ijk}} t = \pd H {\psi^*_{ijk}} \; ,
 \label{ham}
\end{equation}
where
\begin{eqnarray}
H={\frac{1}{2}} \sum_{ijk}&\psi^*_{ijk}&
[-\textstyle{{\frac{1}{12}}}(\psi_{i+2,j,k}-\psi_{i-2,j,k}\nonumber\\
+&\psi_{i,j+2,k}&-\psi_{i,j-2,k}+\psi_{i,j,k+2}-\psi_{i,j,k-2})\nonumber\\
+\textstyle{{\frac{4}{3}}}(&\psi_{i+1,j,k}&-\psi_{i-1,j,k}+\psi_{i,j+1,k}-\psi_{i,j-1,k}\nonumber\\
+&\psi_{i,j,k+1}&-\psi_{i,j,k-1})
]-\textstyle{{\frac{1}{2}}}|\psi_{ijk}|^4
\label{H}
\end{eqnarray}
[in the numerics we set the space step in each direction of the
grid as $dx=dy=dz=1$].

Equation (\ref{ham}) conserves the energy $H$ and the total
particle number $\sum_{ijk}|\psi_{ijk}|^2$ exactly. In time
stepping, the leap-frog scheme was implemented
\begin{equation}
{\rm i}{\frac{\psi_{ijk}^{n+1}-\psi_{ijk}^{n-1}}{2 dt}} = \Big(\pd H {\psi^*_{ijk}}\Big)^n \; ,
 \label{lf}
\end{equation}
with $dt=0.03$. To prevent the even-odd instability of the
leap-frog iterations, we introduce the backward Euler step
\begin{equation}
{\rm i}{\frac{\psi_{ijk}^{n+1}-\psi_{ijk}^{n}}{dt}} = \Big(\pd H {\psi^*_{ijk}}\Big)^{n+1} \; ,
 \label{be}
\end{equation}
every $10^4$ time steps. The leap-frog scheme is nondissipative,
so the only loss of energy and of the total particle number occurs
during the backward Euler step and, since we take this step very
rarely, these losses are insignificant.

The code was tested against known solutions of the GPE:  vortex rings
and rarefaction pulses \cite{jr}. The simulations were performed
on a Sun Enterprise 450 Server and took about three months to
complete for the main set of calculations discussed below.

\subsection{Initial condition}
\label{sec2:b}

To eliminate the  computationally expensive (and the least physically
interesting) transient regime, we started  directly from the
self-similar solution Eq.~(\ref{incond0}) with (\ref{k1})-(\ref{k3}), so that:
\begin{equation}
a_{\bf k} = \sqrt{\xi_{\bf k} n_0 f(\epsilon/\epsilon_0)}
\exp[{\rm i}\phi_{\bf k}],
 \label{ak}
\end{equation}
where $\xi_{\bf k}$ and $\phi_{\bf k}$ are random numbers. [Note
that in the simulations the momentum ${\bf k}$ is the momentum of
the lattice Fourier transform.] The phase $\phi_{\bf k}$ is
uniformly distributed on $[0,2\pi]$ in accordance with the basic
statement of the theory of weak turbulence and with the explicit
microscopic analysis of corresponding quantum field states
\cite{ks2}. The choice of $\xi_{\bf k}$ is rather arbitrary, we
only fix its mean value to be equal to unity, introducing
therefore the parameter $n_0$. The weak-turbulence evolution is
invariant to the details of the statistics of $|a_{\bf k}|$. By
the time the system enters the regime of strong turbulence, the
proper statistics is established automatically since each
harmonic participates in a large number of scattering events. We
tried different distributions for $\xi_{\bf k}$ and saw no
systematic difference in the evolution picture. The main set of
our simulations was done with the distribution function
$w(\xi_{\bf k})=\exp{(-\xi_{\bf k})}$ (heuristically suggested by
equilibrium Hibbs statistics of harmonics in the non-interacting
model).

When choosing the parameters of the initial condition (\ref{incond0})
specified by
the complex Fourier amplitudes (\ref{ak}), we have to take
$\epsilon_0$ small enough to be free from systematic error of
large finite-differences. On the other hand, taking $\epsilon_0$
too small reduces the physical size of the system. Let us define
one period of the amplitude oscillation as $t_p=2\pi/\epsilon_0$
and the number of periods before the blow-up as $P=t_*/t_p$. When
choosing the value of $n_0$ in combination with $\epsilon_0$, we
would like to avoid having $P$ too small when the time scale of
the kinetic regime becomes too short, or having $P$  too large,
when the finite-size effects (the discreteness of the $\bf k$)
dominate the calculation. Given the maximal available grid size
$N=256$, we found that it is optimal to take $n_0=15$ and
$\epsilon_0=1/18$, so that $t_p\approx 113$, $t_*=4\lambda
\pi^3/(\alpha-1)\epsilon_0^2n_0^2 \approx 893$, and $P\approx 8$.
\section{Data processing and results}
\label{sec3}
The instantaneous values of the occupation numbers $n_{\bf
k}(t)=|a_{\bf k}(t)|^2$ are extremely `noisy' functions of time.
To be able to draw some quantitative comparisons and conclusions
we  need either to perform some averaging or to deal with some
coarse-grained self-averaging characteristics of the particle
distribution. Taking the second option, we introduce  {\it shells}
in momentum  space. By the $i$-th shell ($i=1,2,3,\ldots$) we
understand the set of momenta satisfying the condition $i-1 \leq
\log_2(k/2\pi)< i$. The idea behind this definition is that each
shell represents some typical momentum (wavelength) scale and
thus allows to introduce a coarse-grained characteristic of the
occupation numbers corresponding to a given scale. Namely, for
each shell $i$ we introduce the mean occupation number
$\eta_i(t)= \sum^{({\rm shell}~i)} n_{\bf k}(t)/M_i$, where $M_i$
is the number of harmonics in the $i$-th shell. The harmonic ${\bf
k}=0$, which plays a special role (at the very end of the
evolution), is not assigned to any shell.

Another instructive coarse-grained characteristic of the particle
distribution is the integral distribution function $F_{k} =
\sum_{k'\leq k} n_{{\bf k}'}$ which shows how many particles have
momenta not exceeding $k$. We use function $F_{k}$ to keep track of the
formation of the quasi-condensate and to determine wavenumber span
of the above-the-condensate particles. This information is used,
in particular, for filtering out the high-frequency harmonics in
order to interpret the results of our numerical calculations in
superfluid turbulence regime.

With the above-introduced quantities we now turn to the analysis
of the results of  our numerical simulations.
The self-similar character of the evolution
is clearly observed in Fig.~1. Inserts in Fig.~1 give the
comparison of the theoretical prediction of the evolution of the
occupation number function $n_\epsilon(t)$ defined by Eq.~(\ref{k1})
and the evolution of the first and the second shells. The
agreement with the theoretical predictions \cite{svist} is quite
good for $t<600$. After that the numerical solution deviates from
the self-similar theoretical solution which is the manifestation
of the onset of the strong turbulence stage of evolution.

\begin{figure}
\includegraphics[width=\columnwidth]{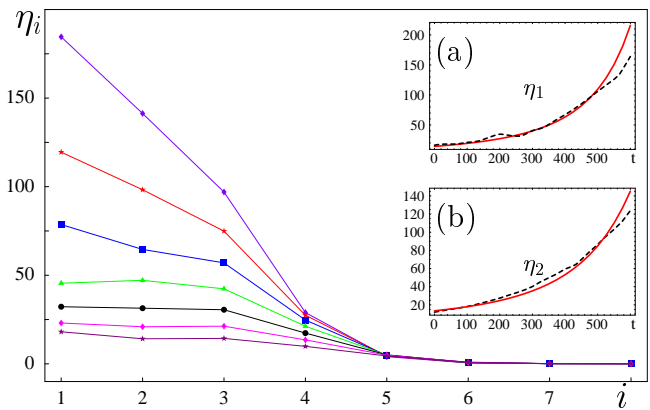}
\caption{The time evolution of $\eta_i(t=100j)$  in the weak
turbulence regime for
$j=0,\ldots ,6$. In the insets we show  the theoretical self-similar solution
(\ref{k1})-(\ref{k3}) (solid line) and the solution obtained
through the  numerical integration of (\ref{gpDimless}) (dashed line)
for  the shells $i=1$ (a) and $i=2$ (b).
}
\end{figure}

As follows from  the dimensional analysis (see, e.g., Ref.~\cite{svist2}),
the characteristic time, $t_0$, and the characteristic wave
vector, $k_0$, at the beginning of the strong turbulence regime
are given by the relations
\begin{equation}
t_*-t_0 \sim C_0 [\hbar^{2\alpha +5}/m^3U^2A^2]^{1/(2\alpha -1)}
\; ,
 \label{t_0}
\end{equation}
\begin{equation}
k_0 \sim C_1 [ A U (m/\hbar)^{\alpha +1}]^{1/(2\alpha -1)} \; ,
 \label{k_0}
\end{equation}
where $C_0$ are $C_1$ some dimensionless constants. Our numerical
results (Fig.~1) indicate that $t_*-t_0 \sim 300$ which implies
$C_0 \sim 40$.  After the formation of the
quasi-condensate ($t>1000$), the distribution of particles
acquires a bimodal shape which is seen in Fig.~2. The salient
characteristic  of the distribution is the shoulder which becomes
sharper and sharper as the evolution continues. Note that by
definition of the function $F_k$, the height of the shoulder is
equal to the number of the quasi-condensate particles. From Fig.~2
we estimate $k_0 \sim 15$ which means that
\begin{equation}
C_1 \sim C_0 \sim 40 \; .
 \label{C_1}
\end{equation}
\begin{figure}
\includegraphics[width=\columnwidth]{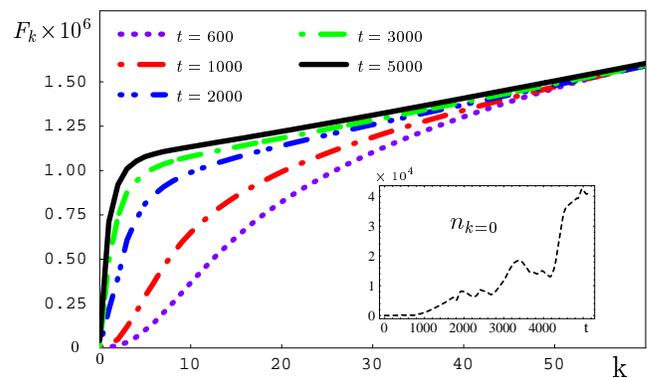}
\caption{Evolution of the integral distribution of particles $F_k=
\sum_{k'\leq k} n_{{\bf k}'}$. Notice the appearance  of a
`shoulder' of $F_k$ indicative of the quasi-condensate formation.
The evolution of $n_{{\bf k }=0}$ is presented in the inset. Note
the strong fluctuations typical for the evolution of a single
harmonic. The fluctuations are also seen in the graph of the first
shell (see insert (a) of Fig.~1).  } \centering
\medskip
\end{figure}
\begin{figure*}[t]
\includegraphics[height=4in]{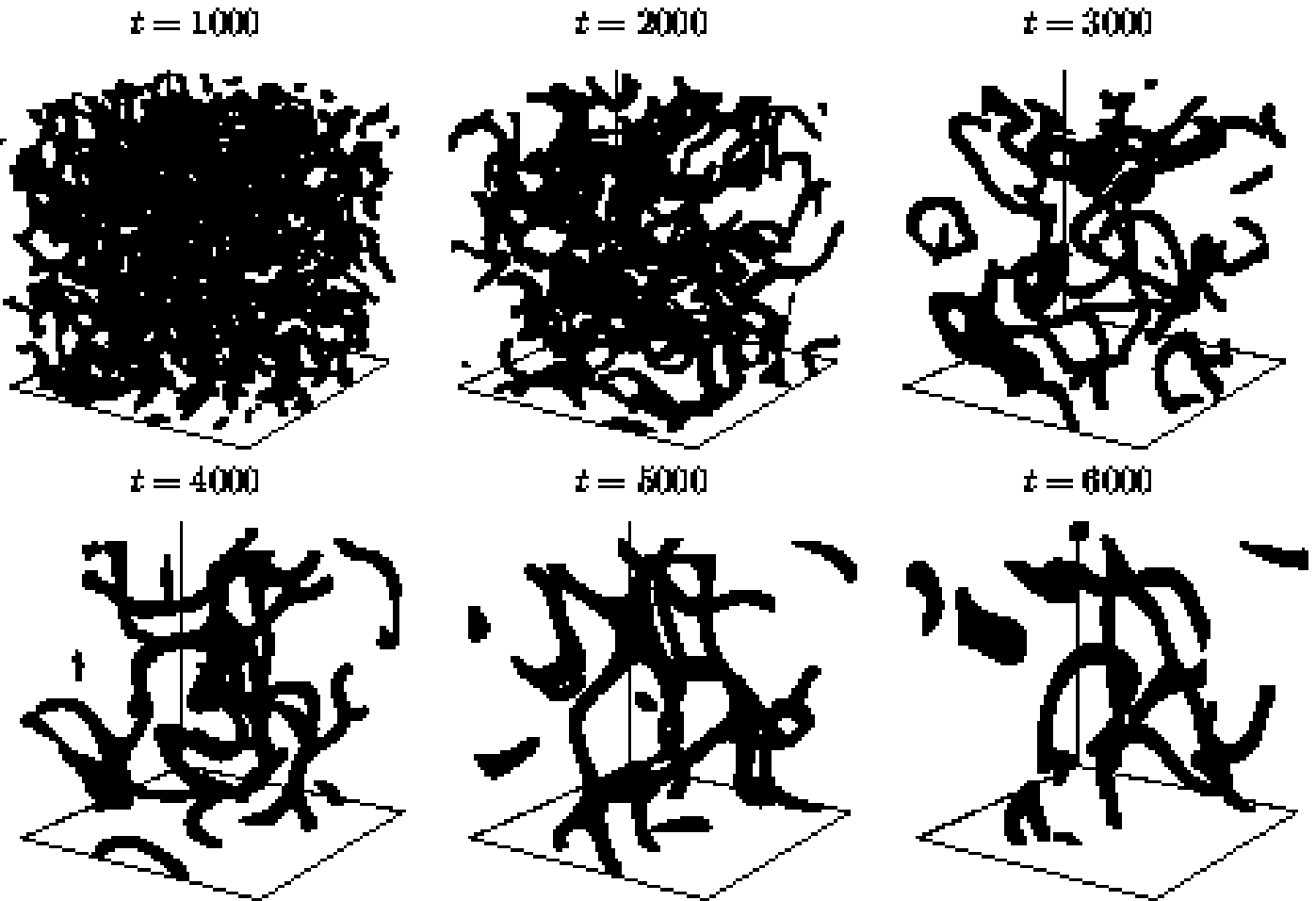}
\caption{Evolution of topological defects in the phase of the
long-wavelength part $\tilde{\psi}$ of the field $\psi$ in the computational box $256^3$. The
defects are visualized by isosurfaces
$|\tilde{\psi}|^2=0.05\langle|\tilde{\psi}|^2\rangle$.
High-frequency spatial waves are suppressed by the factor
$\max\{1-k^2/k_c^2,0\}$, where the cut-off wave number is chosen
according to the phenomenological formula $k_c=9-t/1000$. }
\centering 
\end{figure*}
\begin{figure*}
\includegraphics[height=6 in]{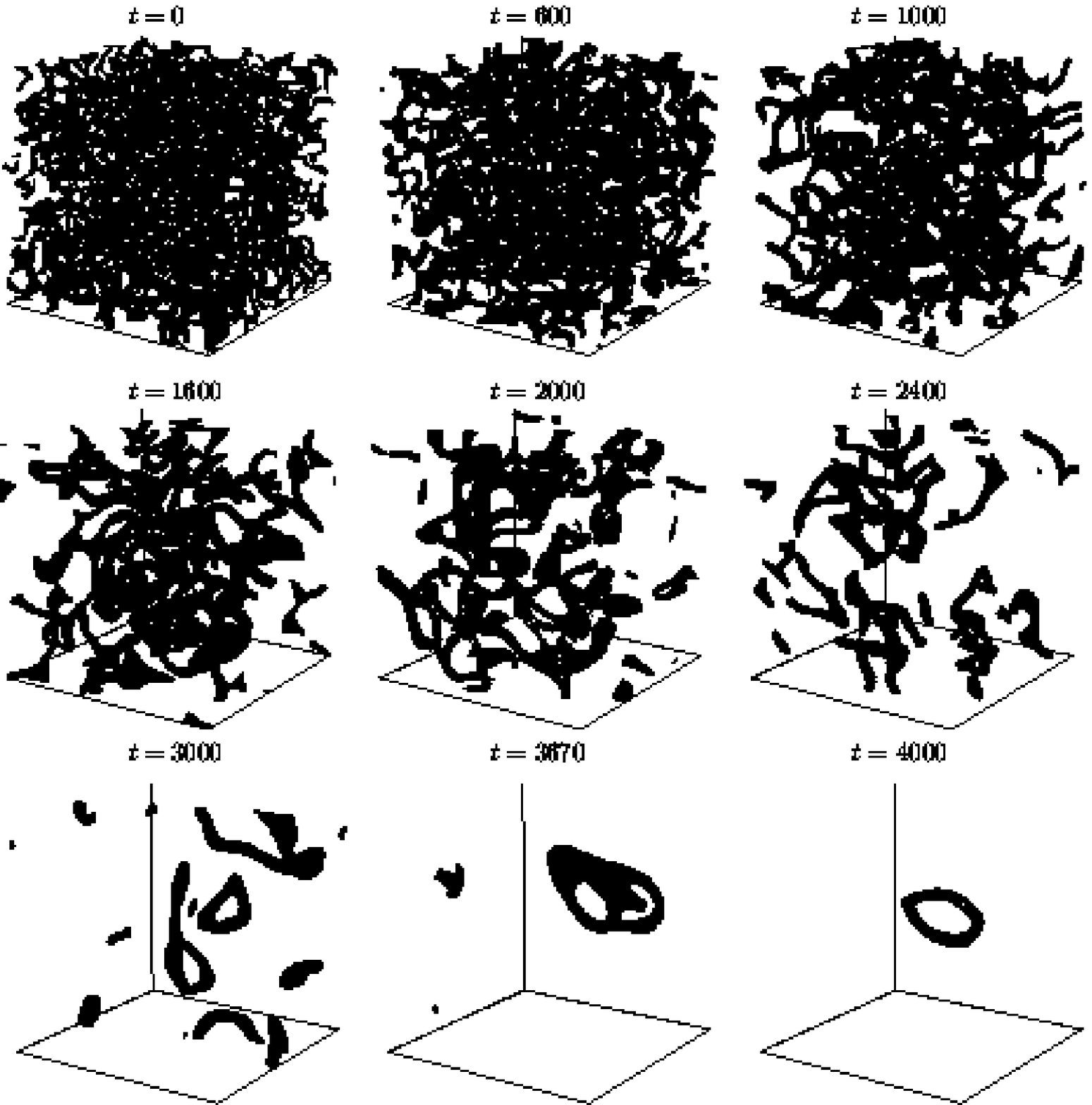}
\caption{Evolution of topological defects in the phase of the
long-wavelength part $\tilde{\psi}$ of the field $\psi$ in the computational box $128^3$. The
defects are visualized by isosurfaces
$|\tilde{\psi}|^2=0.05\langle|\tilde{\psi}|^2\rangle$.
High-frequency spatial waves are suppressed by the factor
$\max\{1-k^2/k_c^2,0\}$, where the cut-off wave number is chosen
according to the phenomenological formula $k_c=9-t/1000$. }
\centering 
\end{figure*}

Within the coherent regime the momentum distribution of the
harmonics yields rather incomplete picture of the evolution and
it becomes reasonable to follow the ordering process in coordinate
space.   The tracing the topological defects in
the phase of the long-wavelength part of the complex matter field
$\psi$ is very important since the transformation of these defects
into a tangle of
well-separated vortex lines is the most essential feature of the
superfluid short-range ordering \cite{ks1}. To this end we first
filter out the high-frequency harmonics by performing the
transformation $a_{\bf k} \rightarrow a_{\bf k} \max\{1 -
k^2/k_c^2,0\}$, where $k_c$ is a cut-off wave number. When the
function $F_k$ has a pronounced quasi-condensate shoulder, the
natural choice is to take $k_c$ somewhat larger than the momentum
of the shoulder in order to remove the above-the-condensate part
of the field $\psi$. In the regime of weak turbulence, when there
is no quasi-condensate, the procedure of filtering is ambiguous:
the distribution is not bimodal, so there is no special low
momentum $k_c$; also, the structure of the defects in the filtered
field essentially depends on the cut-off parameter and, thus, has
no physical meaning.

The results of visualizing the topological defects are presented
in Fig.~3. The formation of a  tangle of well-separated vortices
 and the decay of superfluid
turbulence are clearly seen. This is the key point of our
simulation. To the best of our knowledge, this is the first
unambiguous demonstration of the formation of the state of
superfluid turbulence in the course of self-evolution of weakly
interacting Bose gas. This result forms a solid basis for the
analysis of the further stages of long-range ordering in terms of
well-developed theory of superfluid turbulence that was performed
in Ref.~\onlinecite{ks1} (see also Ref.~\onlinecite{ks3}).

The characteristic time of the evolution of the vortex tangle
depends on the typical interline spacing, $R$, as
$R^2/\ln(R/a_0)$, where $a_0$ is the vortex core size (see, e.g.,
\cite{ks1}). During the final stage of evolution, when $R$ is of
the order of linear size of the computational box, the slowing
down of the relaxation process makes numerical simulation of the
final stage of the vortex tangle decay to be  enormously
expensive in a large computational box.
For example, according to the above-mentioned estimate
of the relaxation time, to achieve the complete disappearance of
the vortex tangle in our $N=256$ system we would need several
years. To observe this final stage of the vortex tangle decay, we
repeated the calculations for a smaller computational box with
$N=128$ (reducing in this way the computational time by a factor
of $\sim 32= 2^3\times 2^2$); see Fig.~4. Parameters of the
initial condition are $\epsilon=1/2$ and $n_0=2 \pi$, so that the
number of periods before the blow-up
 is $P\approx 5$.
A single vortex ring remains at $t=4000$ as the result of the
turbulence decay; see Fig.~5(a).

\begin{figure}
\includegraphics[width=\columnwidth]{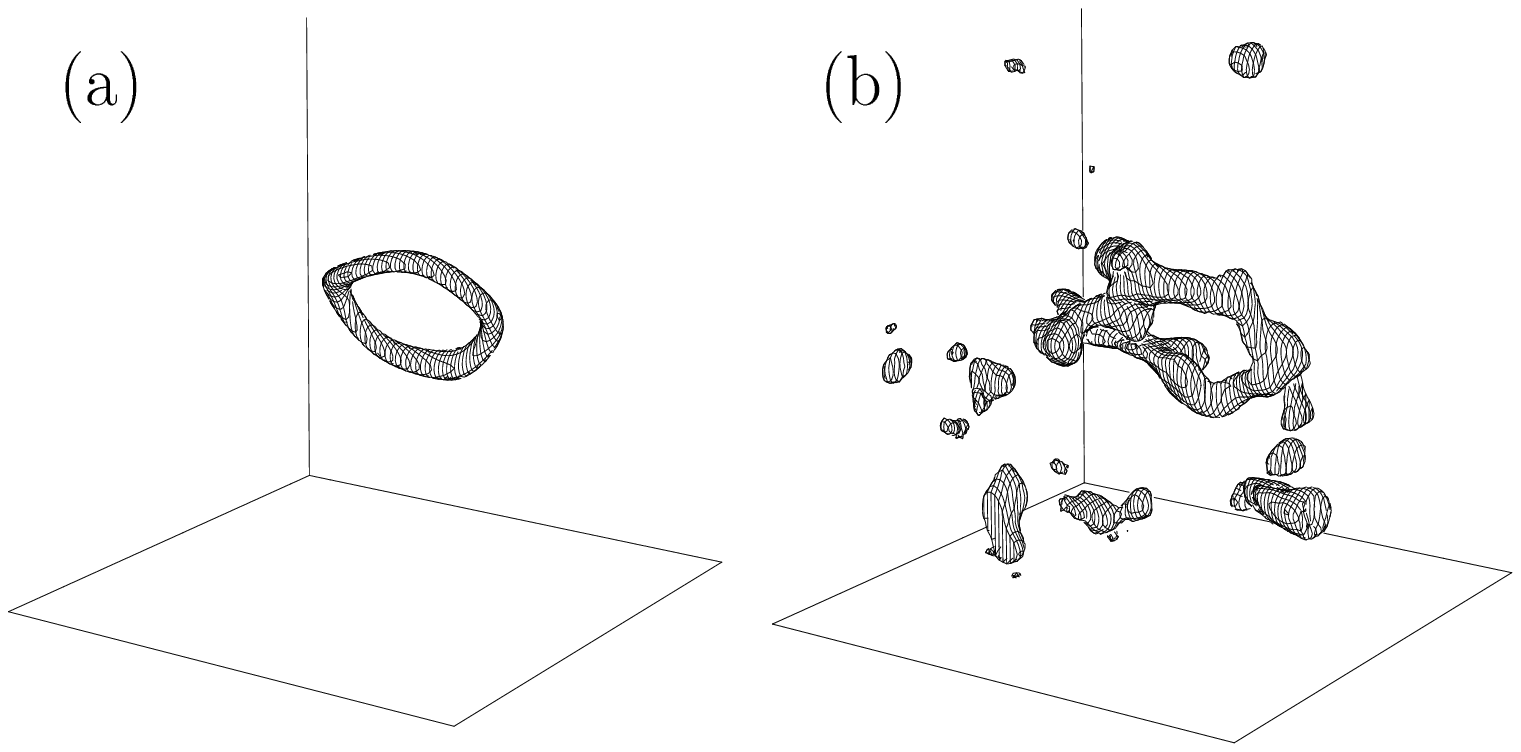}
\caption{Comparison of two isosurfaces obtained by different
filtering techniques. The solution at $t=4000$ is obtained by
numerical integration of (\ref{gpDimless}) in the periodic box
with $N=128$. The isosurface
$|\tilde{\psi}|^2=0.05\langle|\tilde{\psi}|^2\rangle$ is plotted
in Fig.~5(a) using the high-frequency filtering with $k_c=5$. The
isosurface $|\widehat \psi|^2=0.2\langle|\widehat \psi|^2\rangle$
is plotted in Fig.~5(b), where $\widehat \psi$ is defined by
Eq.(\ref{tave}). } 
\end{figure}

The above-mentioned filtering method allows us to visualize the
position of the core of a quantized vortex line, but not the
actual size of the core since we force the solution to be
represented by a relatively small number of harmonics. To get a
better representation of an actual size of the core as well as to
resolve another objects of interest---rarefaction pulses
\cite{jr}, which are likely to appear in the course of
transformation of strong turbulence into superfluid turbulence, we
implement a different type of filtering based on time averaging.
We introduce a Gaussian-weighted time average of the field $\psi$:
\begin{equation}
 \widehat \psi_{ijk}(t) = \int \psi_{ijk}(\tau)
\exp[-(\tau-t)^2/100]\,d\tau \; .
\label{tave}
\end{equation}
The width of the Gaussian kernel in (\ref{tave}) is chosen in such
a way that the (disordered) high-frequency part of the field
$\psi$ is averaged out revealing  the strongly correlated
low-frequency part $\widehat \psi$. Fig.~5 compares the density
isosurfaces obtained by two different methods: by high-frequency
suppression (Fig.~5a) and by time averaging (Fig.~5b). In the
latter case we reveal the actual shape of the vortex core and
resolve the rarefaction pulses.
\section{Conclusion}
\label{sec4}
We have performed large scale numerical simulations of the process
of strongly non-equilibrium Bose-Einstein condensation in a
uniform weakly interacting Bose gas. In the limit of weak
interaction under the condition of strong enough deviation from
equilibrium the key stage of ordering dynamics---superfluid
turbulence formation---is universal and corresponds to the process
of self-ordering of a classical matter field which dynamics is
governed by the time-dependent Gross-Pitaevskii equation
(defocusing nonlinear Schr\"odinger equation). The universality
implies independence of evolution from the details of initial
processes such as, for example, cooling mechanism and rate as
well as from quantum effects such as spontaneous
scattering. All the information about the evolution preceding the
universal stage is absorbed in the single parameter $A$ that
defines scaling of characteristic time and wavenumber, in
accordance with Eqs.~(\ref{t_0})-(\ref{C_1}).

The most important features of the BEC formation scenario observed
in our simulation are as follows. The low-energy part of the
quantum field characterized by large occupation numbers and, thus,
described by a classical complex matter field $\psi$ obeying
Eq.~(\ref{gpDim}) initially evolves in a weak-turbulent
self-similar fashion according to Eqs.~(\ref{k1})-(\ref{k3}). The
occupation numbers at small energies become progressively larger.
At the characteristic time moment $t_0$, given by Eq.~(\ref{t_0}),
close to the formal blow-up time $t_*$ of the solution
(\ref{k1})-(\ref{k3}), the self-similarity of the energy
distribution breaks down. The distribution gradually becomes
bimodal; the low-energy quasi-condensate part of the field sets to
the state of superfluid turbulence characterized by a tangle of
the vortex lines. The further evolution of the quasi-condensate is
independent of the rest of the system (apart from a permanent flux
of the particles into the quasi-condensate) and basically is the
process of relaxation of superfluid turbulence. All vortex lines
relax in a macroscopically large time.

In the present paper, we  dealt with the case of
macroscopically large uniform system. As far as the case of a
trapped gas is concerned, the situation becomes sensitive to the
competition between finite size and nonlinear effects. If
nonlinear effects dominate, the basic physics of the ordering
process is predicted to be analogous to that revealed by our
simulation \cite{svist3}. If finite-size effects dominate (which means
that the initial size of the condensate is smaller than
corresponding healing length, so that, for example, vortices
cannot arise in principle \cite{svist3}), the ordering kinetics is
substantially simplified being reduced to the growth of genuine
condensate \cite{Gardiner,Bijlsma}. Clearly, our numerical approach
can be extended to the case of a trapped Bose gas by simply
including a term with an external potential in Eq.~(\ref{gpDim}).
Such a simulation could provide a deeper interpretation of the
first experiments on the kinetics of BEC formation
\cite{Miesner,Kohl} answering, in particular, the question of
whether the process involves the formation of vortex tangle; and
if not, under which conditions one may expect formation of
superfluid turbulence (quasi condensate) in a realistic
experimental situation.

 \acknowledgments

NGB was supported by the NSF grant DMS-0104288.
BVS acknowledges a support from Russian Foundation for Basic
Research under Grant 01-02-16508 and from the Netherlands
Organization for Scientific Research (NWO).
The authors are very grateful
to Professor Paul Roberts for  fruitful discussions.



\begin{thebibliography}{99}
\bibitem{commentStoof}  In \cite{Stoof} it was argued that the quantum instability
leads to a spontaneous  formation of a homogeneous condensate on
extremely short time scale. The evidence of this instability was
presented using the path-integral version of Keldysh formalism
developed  for the Bose gas. The appearance  of a term that
corresponds to a uniform time-dependent external field in the
effective action and changes its sign was used as the indicator of
this instability. Notice, however, that in the non-Euclidean case
under consideration such a term can be immediately absorbed into
the phase of the field and, therefore, cannot effect the
evolution.
\end{thebibliography}
\end{document}